# Measuring gravity by holding atoms

**Authors:** Cristian D. Panda*[1], Matthew J. Tao[1], Miguel Ceja[1], Holger Müller*[1]

[1]Department of Physics, University of California, Berkeley, 94720, CA, USA.

*Corresponding authors. Email: cpanda@berkeley.edu, hm@berkeley.edu

**Abstract:**

Despite being the dominant force of nature on large scales, gravity remains relatively elusive to experimental measurement. Many questions remain, such as its behavior at small scales or its role in phenomena ascribed to dark matter and dark energy. Atom interferometers are powerful tools for probing gravitation, including Earth's gravity[1,2], the gravitational constant[3], dark energy theories[4–7] and general relativity[8]. However, they typically use atoms in free fall, which limits the measurement time to only a few seconds[9], and to even briefer intervals when measuring the interaction of the atoms with a stationary source mass[3–5,10]. Recently, interferometers with atoms suspended for as long as 70 seconds in an optical lattice have been demonstrated[11–15]. To keep the atoms from falling, however, the optical lattice must apply forces that are billion-fold as strong as the putative signals, so even tiny imperfections reduce sensitivity and generate complex systematic effects. As a result, lattice interferometers have yet to demonstrate precision and accuracy on par with their free fall counterparts and have yet to be used for precision measurement. Here, we optimize the gravitational sensitivity of a lattice interferometer and use a system of signal inversions and switches to suppress and quantify systematic effects. This enables us to measure the attraction of a miniature source mass to be $a_{\mathrm{mass}} = 33.3 \pm 5.6_{\mathrm{stat}} \pm 2.7_{\mathrm{syst}}\,\mathrm{nm/s^2}$, consistent with Newtonian gravity, ruling out the existence of screened dark energy theories[7,16,17] over their natural parameter space (with 95% confidence). More importantly, the combined accuracy of $6.2\,\mathrm{nm/s^2}$ is four times as good as the best similar measurements with freely falling atoms[4,5], demonstrating the advantages of lattice interferometry in fundamental physics measurements. Further upgrades may enable measuring forces at sub-millimeter ranges[18,19], compact gravimetry[20,21], measuring the gravitational Aharonov-Bohm effect[10,22] and the gravitational constant[3], and testing whether the gravitational field itself has quantum properties[23].

Atom interferometry has been instrumental in exploring models of dark energy[4–7]. Cosmological measurements indicate that the universe is expanding at an accelerating rate[24,25], which is consistent with dark energy permeating all of space. These observations could be explained by new scalar fields with screened interactions, which may be present in the vacuum of cosmos and yet hidden from discovery in the high-density environments common to the solar-system and terrestrial physics experiments. Searches for such screened fields, like the chameleon[26,27] or symmetron[28,29], have been performed with neutron interferometry[30,31] or mechanical systems[32–35]. Quantum experiments with atoms in high vacuum near a miniature source mass, however, minimize screening, which gives them favorable sensitivity to dark energy fields over a wide range of masses and coupling constants[4–7]. It is expected that probing the entire remaining chameleon parameter space consistent with the observed cosmic acceleration may be enabled by lattice interferometers with strongly increased sensitivity from prolonged interaction time.[11–15]

In this work, we use a lattice atom interferometer to measure the acceleration $a_{\mathrm{mass}}$ of atoms caused by their interaction (such as gravity) with the source mass. The mass is a hollow tungsten cylinder with height and diameter of 1 inch (Fig. 1a). Each cesium (Cs) atom is in a quantum spatial superposition state, with each interferometer arm held at two lattice sites along the interferometer axis $z$ that are separated by distance $\Delta z$. The atom interferometer measures the potential energy difference, $\Delta U$, between the two arms.

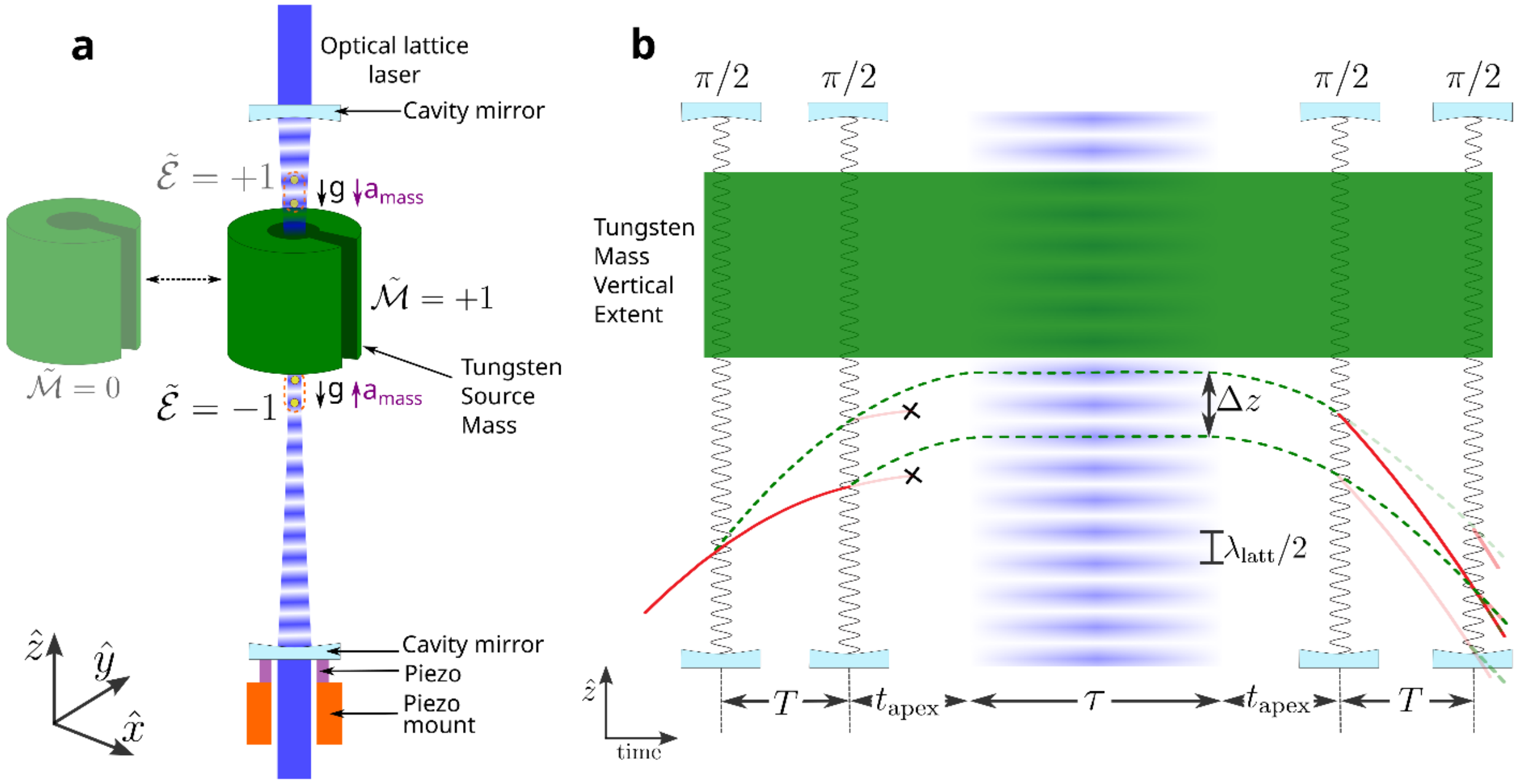

**Figure 1a. Apparatus.** A far-detuned, vertical optical lattice (dark blue, wavelength $\lambda_{\mathrm{latt}} = 943$ nm) is formed by the mode of an optical cavity established by two mirrors (light blue), which is length-stabilized by a ring-piezo (purple). Atoms in a spatial superposition state (yellow circles surrounded by a dashed orange contour) are held in the high-intensity regions of the lattice. They measure the acceleration either above or below the source mass (green). In addition, the source mass can be moved near or far from the atoms. A differential measurement between the $\tilde{\mathcal{E}} \in \{\pm 1\}$ and $\tilde{\mathcal{M}} \in \{+1,0\}$ configurations yields $a_{\mathrm{mass}}$. **b. Trajectories of the atoms shown for the $\tilde{\mathcal{E}} = -1$, $\tilde{\mathcal{M}} = +1$ configuration.** The cavity mode (blue stripes) passes through the center of the tungsten source mass (green). Pairs of $\pi/2$ pulses (wavy vertical lines) separated by time $T$ split, redirect, and interfere the atomic wavepackets. At their apex, the wavepackets are loaded into the optical lattice where they remain for time $\tau$. The internal atomic state is one of the $F = 3$ (red, solid lines) or $F = 4$ (green, dashed lines) hyperfine levels.

The interaction, $a_{\mathrm{mass}}$, is measurable because it contributes a potential difference, $\Delta U_{\mathrm{mass}}$, between the interferometer arms. To isolate $a_{\mathrm{mass}}$ from the ~300 million times larger acceleration due to Earth's gravity, $g$, as well as systematic effects, we use two switches. The first switch reverses the direction of $\boldsymbol{a}_{\mathbf{mass}}$ by positioning the atomic superposition either above ($\tilde{\mathcal{E}} = +1$) or below ($\tilde{\mathcal{E}} = -1$) the source mass. In addition, the source mass can be moved close to ($\tilde{\mathcal{M}} = +1$) or far away from ($\tilde{\mathcal{M}} = 0$) the atoms. Each of these switches not only reject the contributions from $g$ and a wide range of systematic errors, but also help us characterize systematic effects.

The measured phase shift in state $\tilde{\mathcal{M}} \in \{0,1\}$, $\tilde{\mathcal{E}} \in \{-1,1\}$, due to $\boldsymbol{g}$ and $\boldsymbol{a}_{\mathbf{mass}}$ is given by

$$\phi(\tilde{\mathcal{M}}, \tilde{\mathcal{E}}) \approx \frac{\Delta U}{\hbar}\tau = \frac{m_{\mathrm{Cs}}(g + \tilde{\mathcal{M}}\tilde{\mathcal{E}}\, a_{\mathrm{mass}})\, \Delta z}{\hbar}\tau, \tag{1}$$

where $m_{\mathrm{Cs}}$ is the cesium atom mass, $\hbar$ is the reduced Plank constant and $\tau$ is the interferometer hold time. The value of $a_{\mathrm{mass}}$ is extracted from the change in $\phi$ that is correlated with the position of the atoms ($\tilde{\mathcal{E}}$) and position of the source mass ($\tilde{\mathcal{M}}$), that is with the product $\tilde{\mathcal{M}}\tilde{\mathcal{E}}$. By denoting this correlated component as $\phi^{\mathcal{ME}}$, we obtain

$$a_{\mathrm{mass}} \equiv a^{\mathcal{ME}} = \hbar \cdot \phi^{\mathcal{ME}} / (\tau \cdot m_{\mathrm{Cs}} \cdot \Delta z). \tag{2}$$

**Measurement of the interferometer phase**

Atoms are prepared in a magneto-optical trap (MOT) with subsequent polarization gradient cooling (PGC) and Raman sideband cooling (RSC) to produce a 300 nK sample of Cs atoms in the magnetically insensitive $m_F = 0$ state of the ground state hyperfine manifold (see previous paper[11] for details). The atoms are launched upwards with a moving lattice (Fig. 1b). A pair of $\pi/2$ Raman pulses (where atoms are transferred with 50% probability), separated by time $T$, splits the atomic matter-wave four-fold.

We select two wavepackets that are separated vertically by a distance $\Delta z = 2v_r T$, where $v_{\mathrm{r}} = 3.5\ \mathrm{mm/s}$ is the recoil velocity of Cs atoms from 852 nm photons. These wavepackets share the same internal quantum state and external momentum. When they reach the apex, they are adiabatically loaded into the high-intensity regions of a far-detuned optical lattice (wavelength $\lambda_{\mathrm{latt}} = 943$ nm) with a spatial periodicity $\lambda_{\mathrm{latt}}/2$. The optical lattice beam is mode-filtered by an optical cavity[12,36–38]. During the hold, the interferometer wavepackets accumulate the relative phase shift, $\phi$, due to potential difference $\Delta U$ (Eq. 1).

After a hold time $\tau$, the atomic wavepackets are adiabatically unloaded and recombined using a final pair of $\pi/2$ pulses. Their phase difference $\phi$ determines the probabilities $P_{3,4} = [1 \pm C\cos(\phi)]/2$ that the atoms emerge in either the $F = 3$ or $F = 4$ state. The fringe contrast $C$ in the absence of decoherence is $C_0 = 0.5$ because only two of the four interferometer outputs interfere. For detection, we excite the atoms on the Cs D2 line and image the resulting fluorescence signals $S_{3,4}$, which are proportional to $P_{3,4}$. To remove variations in the atom number, both signals are measured simultaneously on the same camera image, using a push beam to spatially separate the $S_{3,4}$ populations (Fig. 2a). From the populations, we then compute the asymmetry,

$$A = (S_3 - S_4)/(S_3 + S_4) = C\cos(\phi). \tag{3}$$

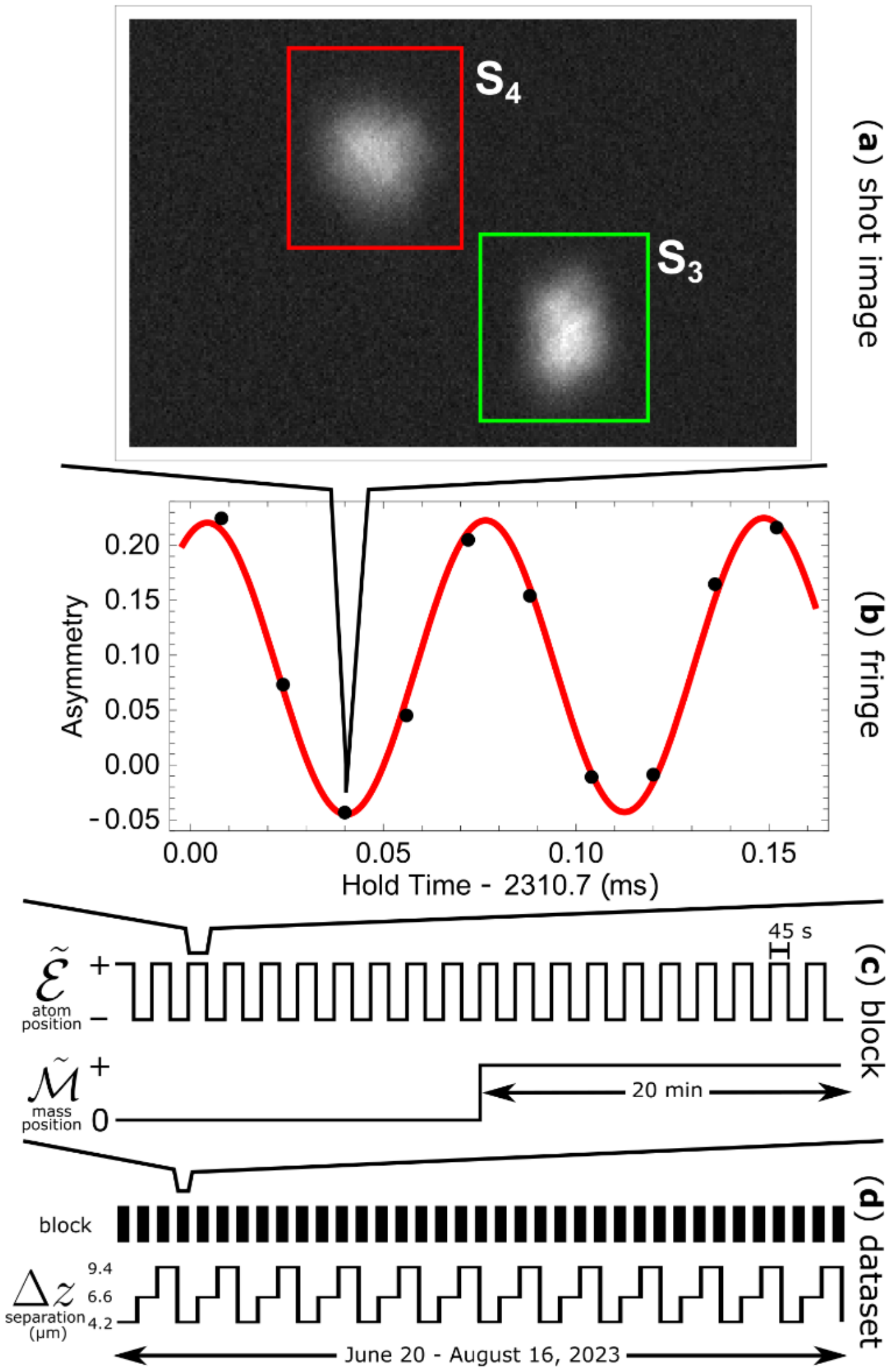

**Figure 2. Experiment timescales. a.** Fluorescence image. $S_4$ and $S_3$ are the signal intensities summed over the red and green squares. **b.** Measured experimental fringe that typically consists of 10 asymmetry points versus hold time $\tau$. **c.** Switches performed within a block. Switch $\tilde{\mathcal{E}}$ alternates from fringe to fringe, while switch $\tilde{\mathcal{M}}$ alternates every 20 fringes. **d.** Dataset measuring $a_{\mathrm{mass}}$ accumulated over about two months (contains 552 'blocks', not all shown). The interferometer separation, $\Delta z$ ($\mu$m), is varied between three values from block to block, over the entire dataset.

We measure $\phi$ by recording $A$ while scanning the hold time $\tau$ in consecutive iterations (Fig. 2b) and fitting the resulting fringe to a sine wave with the phase $\phi$, contrast $C$ and an overall offset as fit parameters.

## Sensitivity, data analysis and statistics

While previous work focused on demonstrating long-lasting coherence[11], here we require high sensitivity to acceleration within a given integration time and therefore a different optimization of the experiment. The theoretical statistical uncertainty at the standard quantum limit (SQL) per experiment shot is given by

$$\delta a_{\mathrm{shot}}^{\mathrm{SQL}} = \hbar/(m_{\mathrm{Cs}} \cdot \Delta z \cdot \tau \cdot C\sqrt{N_{\mathrm{shot}}}), \tag{4}$$

where $N_{\mathrm{shot}}$ is the number of measured atoms. In addition, we empirically determine that contrast decays as $C = C_0 \mathrm{Exp}[-\tau\, \Delta z\, U/(120\ \mu\mathrm{m} \cdot \mathrm{s} \cdot E_{\mathrm{r}})]$ (see reference [11]), where the trap depth $U$ is measured in units of $E_{\mathrm{r}} = m_{\mathrm{Cs}} v_{\mathrm{r}}^2/2 = \hbar \cdot 2\pi \cdot 2.0663$ kHz, the Cs atom recoil energy at 852 nm. The atom number decays as $N = N_0 \mathrm{Exp}[-\tau /(12\text{ s})]$. Given all these constraints, we find that parameters that optimize sensitivity are $\tau = 2.3$ s and $U = 10\ E_\mathrm{r}$.

For the fringe shown in Figure 2b (which is representative of the entire dataset), $C = 0.13$, $\Delta z = 4.2\ \mu$m, and $N_\mathrm{shot} \approx 30{,}000$, the SQL uncertainty (Eq. 4) is $\delta a_\mathrm{shot}^\mathrm{SQL} \approx 2.2 \cdot 10^{-6}\ \mathrm{m/s^2}$. This value is consistent with the measured $\delta a_\mathrm{shot} = 2.6 \cdot 10^{-6}\ \mathrm{m/s^2}$, showing that the sensitivity of our experiment is consistent with the SQL.

Moreover, $\delta a_\mathrm{shot}$ is an order of magnitude smaller than could have been achieved in previous iterations of the apparatus[12], thanks to several improvements, including improved sample preparation, imaging, and an efficient moving-lattice launch (described in detail in [11]). We also implement an atom elevator based on a far detuned moving optical lattice (wavelength $\lambda_\mathrm{latt} = 943$ nm) to shuttle the atoms to various positions along the cavity axis $z$ (such as $\tilde{\mathcal{E}} = \pm 1$).

We switch the atom position ($\tilde{\mathcal{E}}$ switch) from fringe to fringe and the mass position ($\widetilde{\mathcal{M}}$ switch) every 20 fringes. This forms a 'block' of data, which takes ~40 minutes to record (Fig. 2c). Each block therefore contains 10 measurements of $\phi$ for each of the $2^2$ states corresponding to $\{\tilde{\mathcal{E}}, \widetilde{\mathcal{M}}\}$. We average the 10 measurements by weighing them by the uncertainty of each measurement.

We then form 'parity components' [39,40] of the phase, $\phi^{XY}$, which are linear combinations of the measurements that are odd under switch operations $X$ and $Y$ and even under all the other switch operations considered. A superscript 'nr' (for non-reversing) denotes a quantity that is even to all switches. In particular, $a_\mathrm{mass}$ is extracted from $\phi^{\mathcal{M}\mathcal{E}}$, which is odd under the $\widetilde{\mathcal{M}}$ and $\tilde{\mathcal{E}}$ switches

$$\phi^{\mathcal{M}\mathcal{E}} = [\phi(1,1) - \phi(1,-1) - \phi(0,1) + \phi(0,-1)]/2. \tag{5}$$

We then use Eq. 2 to obtain $a_\mathrm{mass}$.

The $a_\mathrm{mass}$ dataset consists of 552 'blocks' that were taken over the duration of about two months (Fig. 2d). To test whether $a_\mathrm{mass}$ depends on wavepacket separation $\Delta z$, we also varied between three values $\Delta z = 4.2,\ 6.6,\ 9.4\ \mu$m during dataset acquisition. We take approximately equal amounts of data at each separation. We find that $a_\mathrm{mass}$ is independent of $\Delta z$ (Figure 3c).

Figures 3a and b show the statistical distribution of $a_\mathrm{mass}$ block data, which is consistent with a normal Gaussian distribution. A chi-squared test yields a reduced $\chi_r^2 = 1.06 \pm 0.04$, which we account for by multiplying the statistical uncertainty of the measurement $\delta a^{\mathcal{M}\mathcal{E}}$ by $\sqrt{\chi_r^2}$. We observe additional excess noise in the channels $\phi^{\mathcal{M}}$, $\phi^{\mathcal{E}}$ and $\phi^\mathrm{nr}$, which are less protected by the $\widetilde{\mathcal{M}}$, $\tilde{\mathcal{E}}$ switches. This shows that these switches eliminate noise and drift in the experiment.

To prevent experimenter bias, we performed a blind analysis by subtracting an unknown offset from $a^{\mathcal{M}\mathcal{E}}$. We revealed this offset only after data collection, statistical data and systematic error analyses were complete.

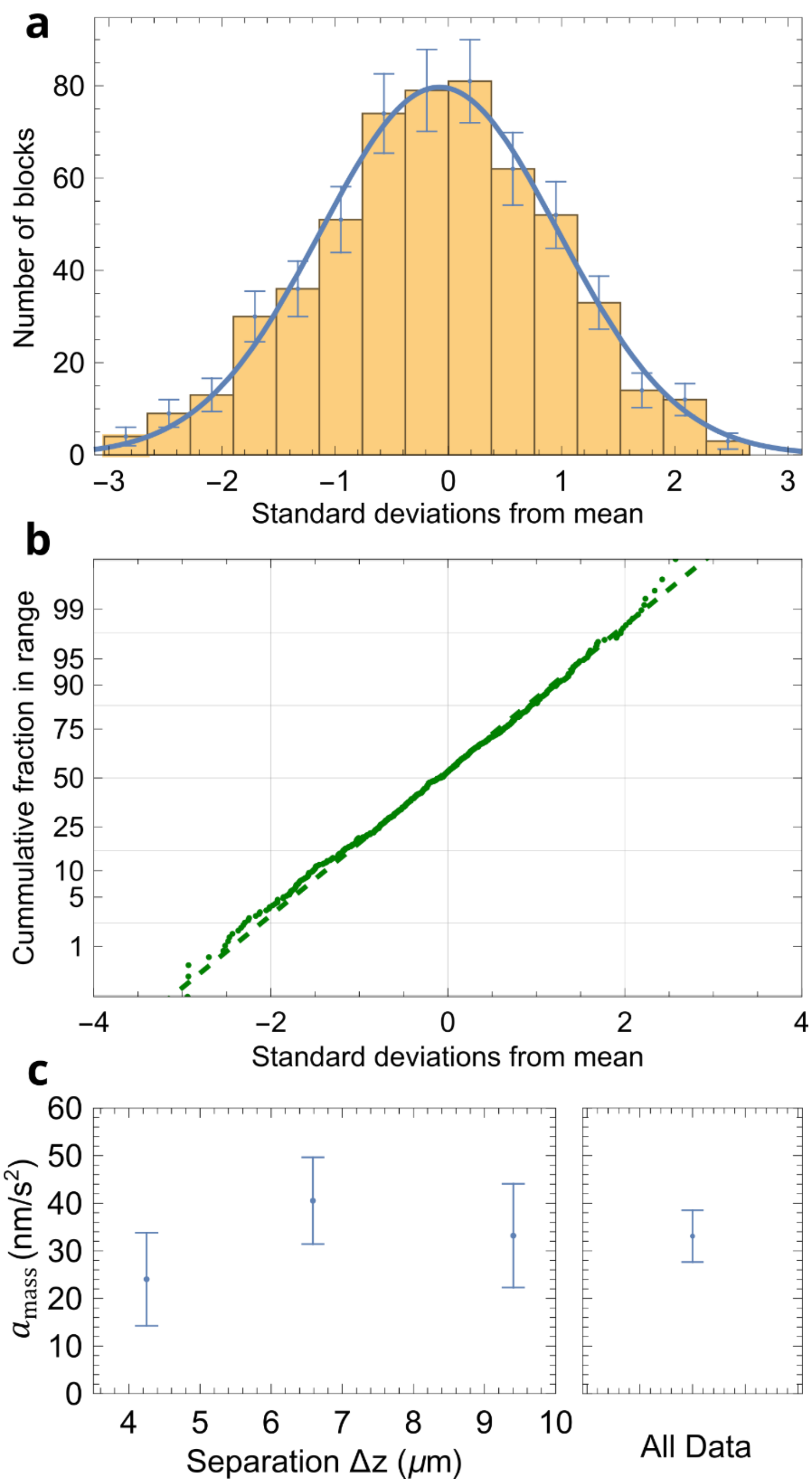

**Figure 3. Statistics of the Measurement Dataset. a. Histogram of centered and normalized $a_{\mathrm{mass}}$ block values.** The values are computed from $(a_{\mathrm{mass}} - \langle a_{\mathrm{mass}} \rangle)/\delta a_{\mathrm{mass}}$, where $\langle a_{\mathrm{mass}} \rangle$ is the average value over the entire dataset. Error bars indicate the standard deviation in the bin expected from a Poisson distribution. The blue line shows a Gaussian fit to the histogram **b. Normal probability plot (green points) compared with a normal distribution (green dashed line).** The vertical axis is scaled such that a Gaussian distribution appears linear. **c. Values of $a_{\mathrm{mass}}$ grouped according to separation, $\Delta z$, and combined for the entire dataset.** Error bars correspond to $1\sigma$ (68% confidence interval).

**Systematic errors**

As described above, we acquire repeated interferometer measurements under varying experimental conditions to (a) isolate the source mass acceleration, $a^{\mathcal{ME}}$, from other background noise and errors and (b) search for possible systematic errors. Since $a^{\mathcal{ME}}$ is the acceleration component that is correlated with both $\widetilde{\mathcal{M}}$ and $\tilde{\mathcal{E}}$ switches, each independently suppresses possible systematic influences of many experimental parameters $P$ on $a^{\mathcal{ME}}$. The uncorrelated parameters, $P^{\mathrm{nr}}$, are suppressed by both switches, while parameters correlated with only one switch, $P^{\mathcal{M}}$ and $P^{\mathcal{E}}$, are still suppressed by the other switch.

To search for sources of systematic error, we vary experimental parameters $P$ over a larger range than typically found in the experiment and measured their influence on $a^{\mathcal{ME}}$. If $P$ was measured (or is theoretically expected) to have a non-zero influence on $a^{\mathcal{ME}}$, we use additional measurements and modeling to determine the systematic dependence of $a^{\mathcal{ME}}$ on $P$, $a^{\mathcal{ME}}(P)$. We use a separated auxiliary measurement to determine the time-averaged ambient value of $P$, $\langle P\rangle$, and then compute the associated systematic shift, $a_P^{\mathcal{ME}}(\langle P\rangle)$. This data was used only for the determination of systematic shifts and uncertainties and is not included otherwise in the measurement dataset.

The only parameter for which a nonzero shift was either observed or expected is <u>blackbody radiation</u>, which is known in our setup to generate forces on the atoms that are given by $a_{\mathrm{BBR}}^{\mathcal{ME}} = -4.3 \pm 0.6 \cdot 10^{-8}\,(T_{\mathrm{mass}}^4 - T_0^4)\ \mathrm{nm}/(\mathrm{K}^4\ \mathrm{s}^2)$, where $T_{\mathrm{mass}}$ is temperature of the source mass and $T_0$ is the temperature of the environment[41]. We use an infrared thermal sensor to measure $T_{\mathrm{mass}}$ and $T_0$, which we find to be equal to within $0.05 \pm 0.3$ K. We use this measurement to compute a shift and systematic uncertainty that are included in the systematic error budget (Table I).

Other parameters $P$ are neither observed nor expected to significantly affect $a^{\mathcal{ME}}$, but are nevertheless included in the error budget, as described below.

<u>AC Stark shift difference between upper and lower atom positions, $a^{\mathcal{E}}$.</u> In the fully retracted position ($\widetilde{\mathcal{M}} = 0$), the mass should cause no measurable difference ($< 0.01\,\mathrm{nm/s^2}$, see Supplementary Information) between the acceleration in the upper ($\tilde{\mathcal{E}} = +1$) and lower ($\tilde{\mathcal{E}} = -1$) positions of the atoms. In the experiment, however, we measure a significantly non-zero average $a^{\mathcal{E}}$ in the final dataset, $\langle P\rangle = \langle a^{\mathcal{E}}\rangle = -377\ \mathrm{nm/s^2}$ with uncertainty $\delta\langle P\rangle = 9\ \mathrm{nm/s^2}$. This value is consistent with a model based on the light-shift (AC Stark shift) between the two elevator ($\tilde{\mathcal{E}} = \pm 1$) positions due to the divergence of the optical lattice mode, as described in detail in Supplementary Material.

Ideally, any effect of $a^{\mathcal{E}}$ on $a^{\mathcal{ME}}$ should be cancelled by the $\widetilde{\mathcal{M}}$ (mass position) switch. To quantify the possible residual influence ("leakage") from $a^{\mathcal{E}}$ to $a^{\mathcal{ME}}$, we generate a large artificial $a^{\mathcal{E}}$ by applying a magnetic field gradient, $\partial B_z/\partial z$. We assume a linear relationship between $P$ and $a^{\mathcal{ME}}$ and use this data to determine the slope, $S_P = \delta a^{\mathcal{ME}}/\delta P$, which we measure to be $S_{a^{\mathcal{E}}} = \partial a^{\mathcal{ME}}/\partial a^{\mathcal{E}} = 2.6 \cdot 10^{-4}$ with an uncertainty $\delta S_P$ of $\delta S_{a^{\mathcal{E}}} = 1.9 \cdot 10^{-4}$. Since $S_P$ is consistent with zero, as expected, we apply no systematic correction but use the measured $S_P$, $\delta S_P$, $\langle P\rangle$ and $\delta\langle P\rangle$ to determine the error bar from

$$\delta a_P^{\mathcal{ME}} = \sqrt{(S_P \cdot \delta\langle P\rangle)\verb|^|2 + (\delta S_P \cdot \langle P\rangle)\verb|^|2}\ . \qquad (6)$$

We include this error bar in the systematic error budget (Table I).

Contributions due to $a^{\mathcal{E}}$ and $\mathcal{M}$- correlated parameters. Additional leakage of $a^{\mathcal{E}}$ into $a^{\mathcal{ME}}$ could result from another parameter that is correlated with the position of the source mass, $P^{\mathcal{M}}$. We identify four such parameters: MOT position, lattice intensity, as well as axial and transverse magnetic fields. We determine the possible systematic error contributions by measuring their associated slopes: $S_{P^{\mathcal{M}}} = \partial a^{\mathcal{ME}}/\partial P^{\mathcal{M}}$, which were all found to be consistent with zero (Table S1). We use $S_{P^{\mathcal{M}}}$ and $\langle P^{\mathcal{M}}\rangle$ for each of the four parameters to calculate limits that we include in the systematic error budget using Eq. 6 (Table I). We discuss each parameter in more detail in the following.

When the mass is inserted ($\widetilde{\mathcal{M}} = 0 \rightarrow +1$), we observe a change in the MOT position at the level of 10 $\mu$m, which is due to the source mass mounting rod partially blocking one of the six MOT laser beams. However, we find that the position of the atoms during the measurement is determined by the cavity mode and therefore largely unaffected by the source mass position. This explains why there is no observed influence of the MOT position on $a^{\mathcal{ME}}$.

Clipping of the cavity laser beam by the source mass is expected to be negligible, as the inner diameter is more than 20 times larger than the radius of the cavity mode. We use the transmission photodetector to observe the intensity of the lattice laser in the $\widetilde{\mathcal{M}} = \{0,1\}$ positions and measure $\langle U^{\mathcal{M}}\rangle$ consistent with zero at the 2 parts in $10^4$ level.

Ferromagnetic impurities may give rise to a magnetization of the source mass. We use an auxiliary measurement to determine the residual magnetic field difference between the $\widetilde{\mathcal{M}} = \{0,1\}$ positions, $\langle B^{\mathcal{M}}\rangle$ to be consistent with zero and smaller than 1 mGauss (see reference [4]). We place independent systematic contributions due to axial (along $z$) and transverse (along $x$, $y$) magnetic fields since they have different effects on the interferometer phase.

Source mass surface. The source mass is electrically grounded. However, thin films of surface oxidation may form an insulating layer, allowing surface voltages of up to 10 V to form. Using the ground state polarizability of cesium[42], even these worst-case scenario voltages would cause a maximum acceleration of only 0.5 nm/s^2. We include this contribution in the systematic error budget (Table I). Casimir–Polder effects are negligible[43], since the atoms never come closer to the source-mass surface than about 4 mm.

In addition to effects above, we varied over 35 additional experimental parameters and measured their effect on $a^{\mathcal{ME}}$ (Table S1). None of these were observed or expected to have an influence on $a^{\mathcal{ME}}$ and therefore corresponding error bars were not included in the systematic error budget.

| Parameter | Shift ($\mathrm{nm/s^2}$) | Uncertainty ($\mathrm{nm/s^2}$) |
|---|---|---|
| Black-body radiation gradient | 0.05 | 1.30 |
| $a^{\mathcal{E}}$(via $\partial B/\partial z$) | | 0.07 |
| $\mathcal{M}$-correlated MOT position | | 1.86 |
| $\mathcal{M}$-correlated trap depth | | 0.31 |
| $\mathcal{M}$-correlated axial B-field | | 0.92 |
| $\mathcal{M}$-correlated transverse B-field | | 0.84 |
| DC Stark Shift | | 0.50 |
| Total systematic | 0.05 | 2.66 |
| Statistical uncertainty | | 5.61 |
| Total uncertainty | | 6.21 |
| Source-mass calculated gravity | 35.20 | 1.00 |

**Table *1*. Systematic shifts and uncertainties in $\boldsymbol{a_{\mathrm{mass}}}$.** All uncertainties are added in quadrature.

## Result and conclusions

After unblinding, we find $a_{\mathrm{mass}} = 33.3 \pm 5.6_{\mathrm{stat}} \pm 2.7_{\mathrm{syst}}\,\mathrm{nm/s^2} = 33.3 \pm 6.2\,\mathrm{nm/s^2}$ for the acceleration of the atoms towards the source mass. The expected acceleration is $a_{\mathrm{mass}}^{\mathrm{calc}} = 35.2 \pm 1.0\,\mathrm{nm/s^2}$ (see Supplementary Information). The difference $a_{\mathrm{anomaly}} \equiv a_{\mathrm{mass}} - a_{\mathrm{mass}}^{\mathrm{calc}} = -1.9 \pm 6.3\,\mathrm{nm/s^2}$ is consistent with zero. The combined statistical and systematic uncertainty of this measurement has been reduced fourfold from the previous best atom interferometric measurements of the gravity due to a cm sized source mass[4,5]. An upper limit $|a_{\mathrm{anomaly}}| < 13\,\mathrm{nm/s^2}$ is computed using a folded Gaussian at 95% confidence, which represents a factor of 6 improvement over the previous results achieved with interferometers where atoms are in free-fall[4,5].

Our measurement improves on previous constraints on fifth forces due to chameleon or symmetron particles[7,16,17] by factors of 3-5. Figure 4 shows the excluded parameter ranges for these models. The available parameter space for chameleons with $\Lambda \approx 2.4$ meV (black line), the dark energy level required to drive cosmic acceleration today, is now fully excluded (Figure 4a). Significant regions of parameter space with the power index describing the shape of the chameleon potential $n > 1$ have also been constrained (Figure 4b). Similar improvements are seen for symmetrons (Figure 4c).

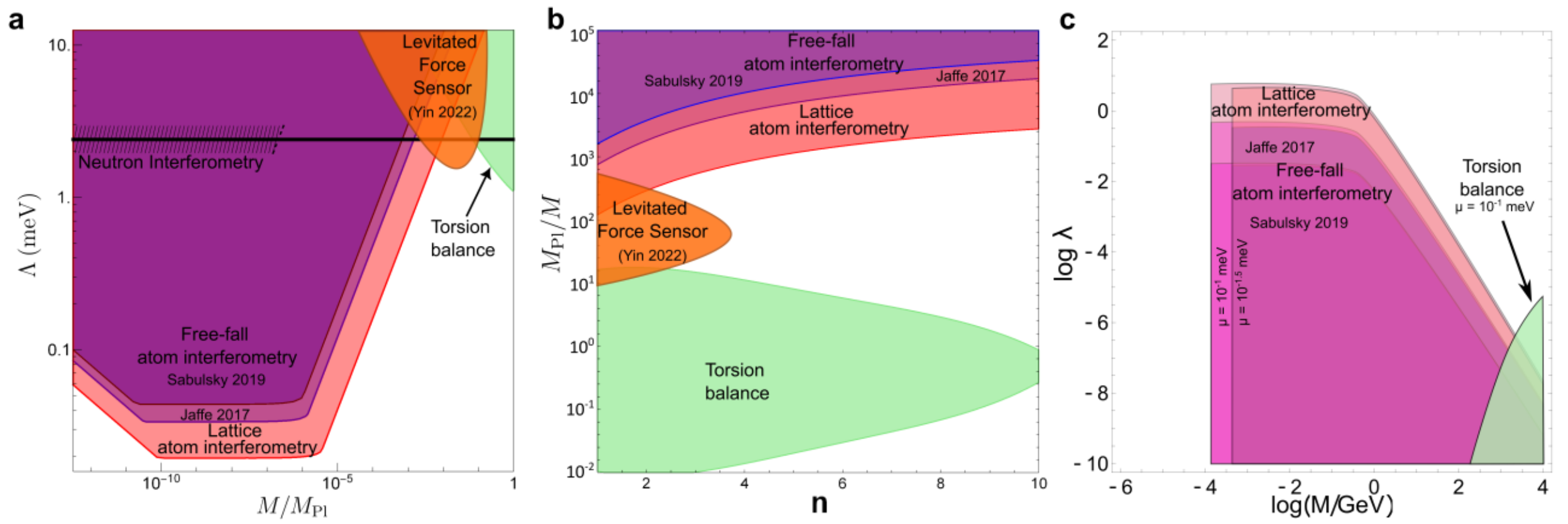

**Figure 4. Constraints on chameleon and symmetron dark energy fields. a. Chameleon fields.** Shaded areas in the $M-\Lambda$ parameter plane of chameleon field are ruled out (see Ref [4] for definitions). $\Lambda \approx 2.4$ meV (black line) is the dark energy level required to drive cosmic acceleration today. Limits from previous experiments are shown: interferometry with atoms in free-fall[4,5], neutron interferometry[30,31], levitated force sensors[32], and torsion balances[33,34]. **b. Chameleon limits for n>1.** Bounds with $\Lambda \approx 2.4$ meV showing the narrowing gap in which chameleon gap remains viable. n is the power law index describing the shape of the chameleon potential. **c. Symmetron fields.** Constraints from atom interferometers and torsion balance experiments are shown. All shaded areas are ruled out at 95% confidence level.

Since their first demonstration more than 30 years ago[2], interferometers with atoms in free-fall have been at the forefront of fundamental science, through measurements of the gravitational constant[3], fine structure constant[44,45], fundamental tests of gravity[8] or searches for new physics[4,5]. We demonstrate the first interferometric measurement of gravity with atoms held by an optical lattice with precision that now surpasses free-fall atom interferometric measurements. This technique enables future applications in the fields of inertial sensing[21,46], by performing gravity maps with high spatial resolution[47] and introducing portable gravity sensors[48]; and fundamental physics, by observing a phase shift in the absence of forces[10,22] or signals from non-classical gravity[18,23].

## Acknowledgments:

We thank A. Reynoso and J. Egelhoff for experimental assistance; J. Lopez, T. Gutierrez and G. Long for technical support; G. Louie and P. Bhattacharyya for discussions and comments on the manuscript; J. Axelrod, B. Elder, M. Jaffe, J. Khoury, P. Haslinger, Y. Murakami, A. Singh and V. Xu and the entire Muller group for valuable discussions.

## Methods

Determination of source mass Newtonian gravitation attraction

We use a combination of analytics, finite element analysis modeling, and spatial triangulation to determine the expected Newtonian gravitational acceleration from the source mass, $a_{\mathrm{mass}}^{\mathrm{calc}}$.

The tungsten source mass is a hollow cylinder with a height of 25.4 mm, outer diameter of 25.4 mm and inner diameter of 10.0 mm. A rectangular slot with width of 5.7 mm allows for insertion and removal of the source mass without blocking the cavity mode. The mass is manufactured using wire electron discharge machining (EDM) with tolerances better than $10\,\mu$m. The calculated source mass volume is consistent with its measured weight given the density of tungsten to within <1%.

To determine the source mass position relative to the atoms, we record sequential images of the atom sample and source mass (Figure S1) using three different camera positions. Measuring the position of the atom sample at three different heights along the atomic elevator axis (which coincides with the cavity axis) fully determines (through triangulation) the orientation of the elevator axis with respect to the source mass. This procedure provides a measurement of the two elevator atom positions ($\tilde{\mathcal{E}} = \pm 1$) with respect to the source mass with better than 1 mm accuracy.

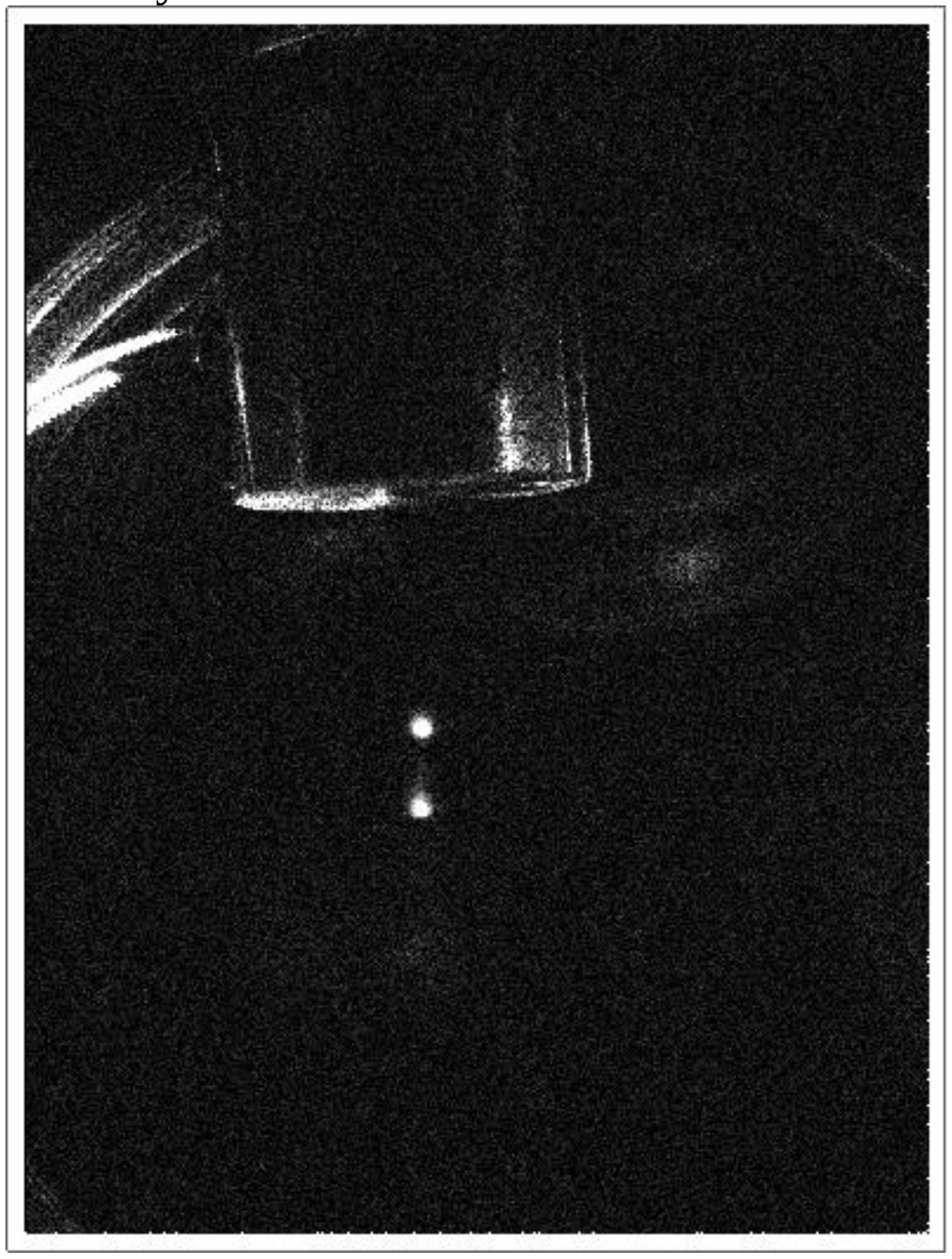

**Figure *S1*. Atom sample entering source mass.** Sequence of images showing the Cs atom sample at various positions along its atomic elevator trajectory. Acquiring this movie from three different perspectives triangulates the position of the atom sample with respect to the source mass with an accuracy better than 1 mm.

To estimate the gravitational acceleration $a_{\mathrm{mass}}^{\mathrm{calc}}$ at this position, we first analytically calculate the gravitational field along the axis of a simple hollow cylinder, disregarding the existence of the slot. We use this calculation to verify the results of a finite element analysis software (COMSOL Multiphysics; since COMSOL does not offer a gravitational module by default, we use the electrostatic module, modifying the "charge" of the source-mass to the density of tungsten and using the gravitational constant instead of electrostatic constant). We find good agreement to better than 0.1%. We then add the rectangular slot to the finite-element model to generate a three-dimensional map of the gravitational field (Figure S2).

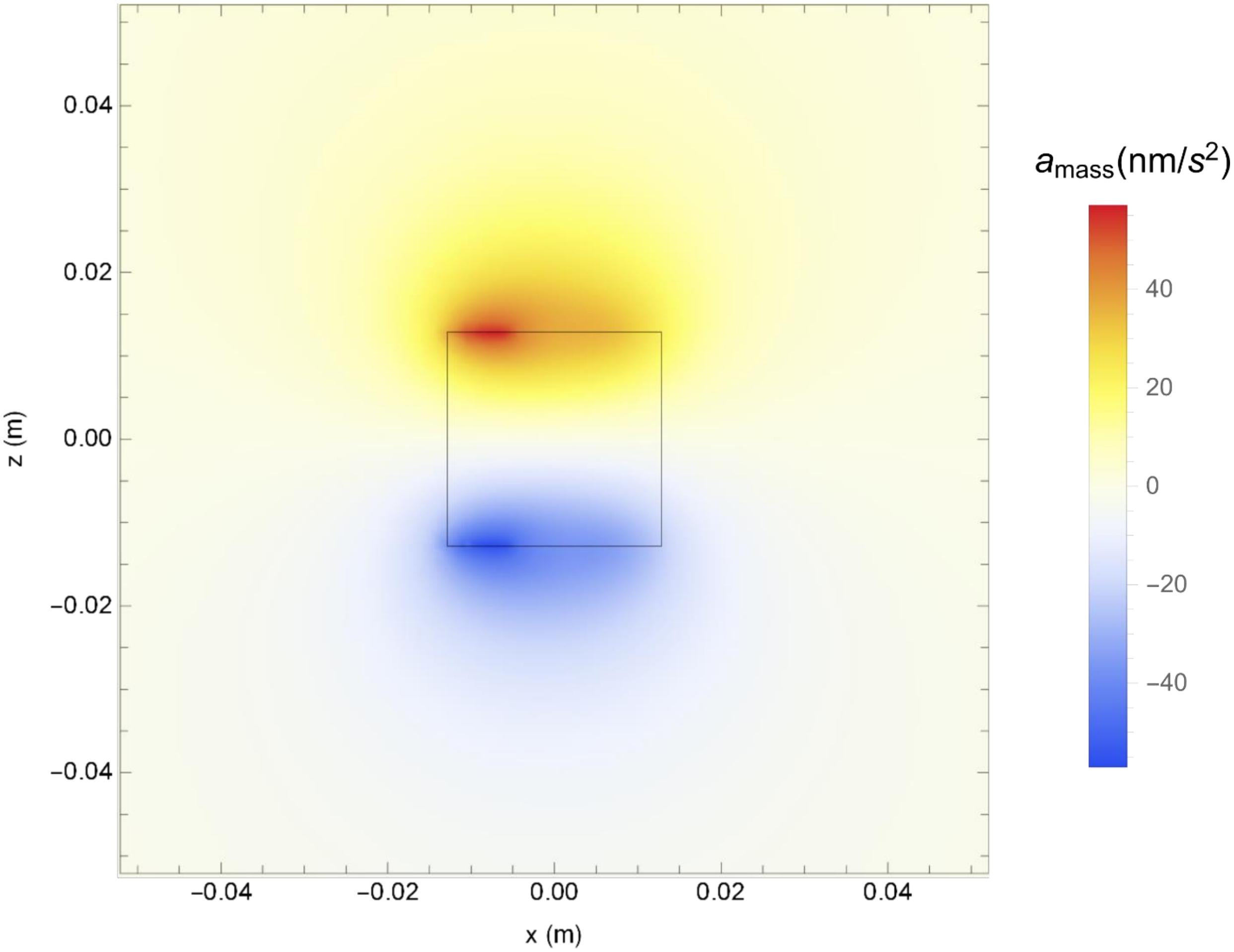

**Figure *S2*. Map of source mass gravity.** A 2D slice of the $z$ component of the gravitational field calculated using fine element analysis in COMSOL is shown. The black square shows the extent of the hollow cylinder. Gravity is stronger on the left side of the map due to the presence of the rectangular slot on the right side.

At the triangulated position, we find the source mass gravitational acceleration along the interferometer axis to equal $a_{\text{mass}}^{\text{calc}} \equiv (a_{\text{mass}}^{\mathcal{E}+} + a_{\text{mass}}^{\mathcal{E}-})/2 = 35.2 \pm 1.0\,\text{nm/s}^2$. The acceleration in the mass-out position ($\widetilde{\mathcal{M}} = 0$) is $< 0.01\,\text{nm/s}^2$ and therefore negligible.

Systematic investigation: $a^{\mathcal{E}}$phase shift model

We describe here in detail our investigations into the mechanism causing the shift in the $a^{\mathcal{E}}$ channel described in the main text. We identify the primary contribution to $a^{\mathcal{E}}$ as a light-shift (AC Stark shift) that is differential between the two interferometer arms, $a^{\text{ls}}$. It differs between the two elevator positions ($\tilde{\mathcal{E}} = \pm 1$) and varies linearly with $z$ due to the divergence of the optical lattice mode.

Modelling the lattice laser beam as a Gaussian beam, its intensity varies as $I(z) = I_0[1 - (\lambda\, z)^2/(\pi w_0^2)^2]$, where $w_0 = 760\,\mu\text{m}$ is the waist and $z$ is the vertical position with respect to the waist. At each vertical interferometer position, $z$, the difference between the intensity of the two interferometer arms is given by

$$\Delta I(z) = \partial I/\partial z \cdot \Delta z = -\,2I_0\lambda^2/(\pi w_0^2)^2\, z\,\Delta z.$$

Using Eq. 1, the measured acceleration is given by

$$a^{\mathrm{ls}}(z) = \Delta U/m_{\mathrm{Cs}}\Delta z \;= -2\,U_0\lambda^2/(\pi w_0^2)^2\, z\,/m_{\mathrm{Cs}}.$$

This results in a differential acceleration shift during usual data-taking of $a^{\mathcal{E}} = (a^{\mathrm{ls}}(z^{\mathcal{E}+}) - a^{\mathrm{ls}}(z^{\mathcal{E}-}))/2$, where $z^{\mathcal{E}\pm}$ are the vertical positions of the atoms at the two elevator positions.

To verify this model, we took an auxiliary dataset to measure $a^{\mathrm{ls}}(z) - a^{\mathrm{ls}}(0)$ at various $z$ positions along the lattice axis in an auxiliary measurement (Fig. S3).

Since the above model unrealistically assumes atoms at zero temperature, we estimate $a^{\mathrm{ls}}(z)$ based on simulations of the trajectories of the atoms inside the optical lattice at the observed temperature of 300 $\mu$K, as described in[1]. Both the analytical model and the simulation are found to be in good agreement with the main dataset $\langle a^{\mathcal{E}}\rangle$ (Fig. S2). The model was further confirmed by our observation of a linear scaling of $a^{\mathcal{E}}$ with the trap depth $U_0$.

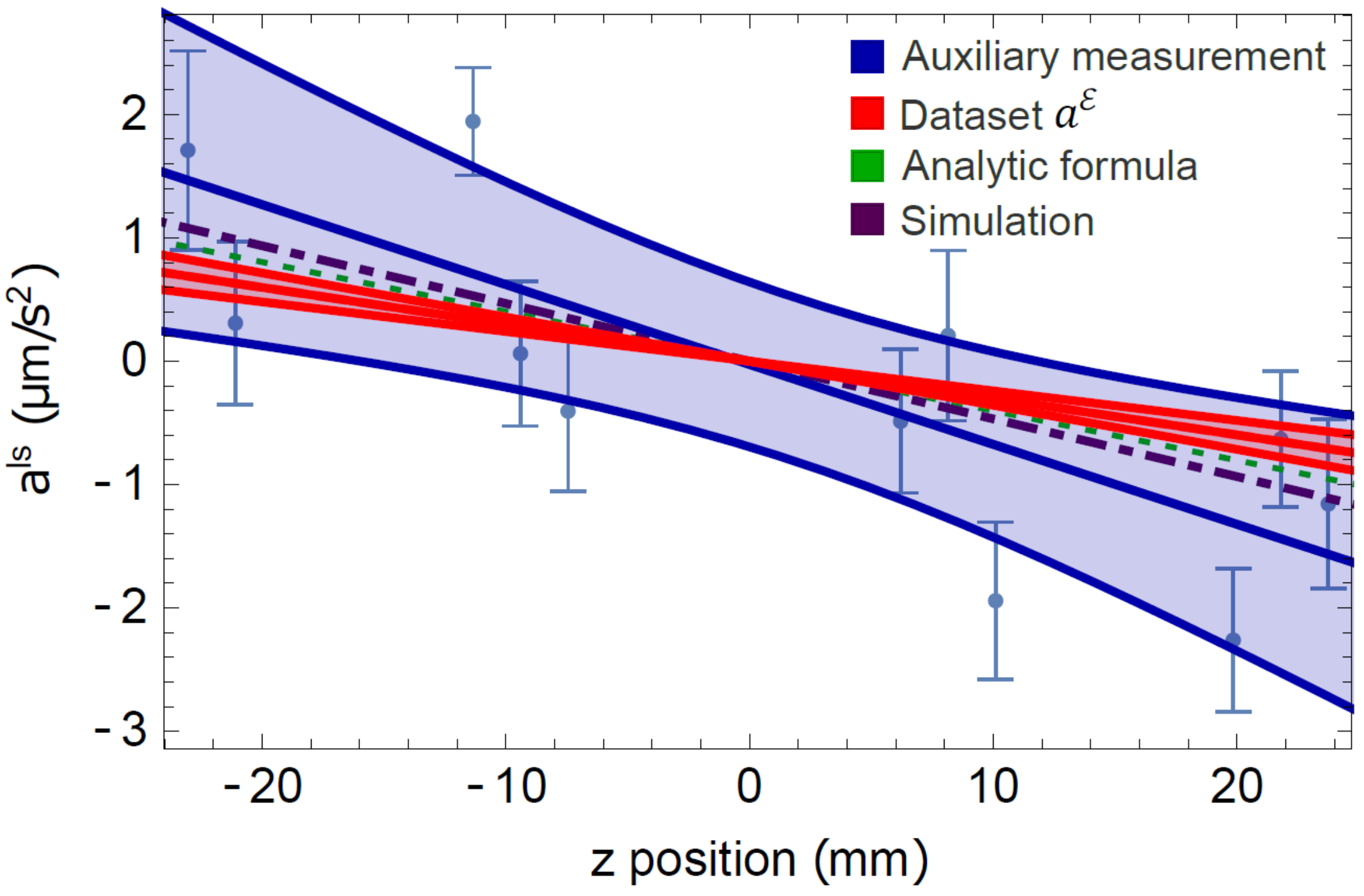

**Figure *S3*. Acceleration shift due to lattice divergence $a^{\mathrm{ls}}$ .** In an auxiliary measurement, we observe a linear change in measured acceleration $a^{\mathrm{ls}}$ as a function of vertical position $z$. This is due to the differential AC Stark shift from the changing trap potential, $\Delta U(z)$, as the atoms are held in various positions along the diverging lattice potential. We observe agreement between the analytic equation derived above, simulation, and experiment within experimental uncertainty. The bands correspond to 95% (2 sigma) confidence intervals.

Parameters varied in the search for systematic errors

Table S1 shows a list of parameters that were varied while searching for unexpected systematic errors. The procedure for performing these checks is described in the main text.

| Category | Parameter Varied | Unit | Applied value(s) (unit) | Ambient variation, $\delta\langle P\rangle$ (unit) | Slope Mean, $S_P$ $(\mathrm{nm/s^2/unit})$ | Slope Uncertainty, $\delta S_P$ $(\mathrm{nm/s^2/unit})$ |
|---|---|---|---|---|---|---|
| **Lattice parameters** | Trap depth | arb unit | 0.9, 1.2, 1.6 | 0.01 | 165 | 180 |
| | Separation | $\mu$m | $4.2, 6.6, 9.4$ | 0.05 | 74 | 133 |
| | Hold time | s | 1.5, 2.2, 2.8, 3.6 | $10^{-6}$ | 0.0179 | 0.0581 |
| | Lattice laser polarization ellipticity | % | 0,40 | 1 | -0.62 | 2.37 |
| | Lattice laser frequency noise | arb unit | 1,20 | 1 | 7.6 | 4.6 |
| | Transverse temperature (via LG10 mode) | mK | 0.3,0.16 | 0.001 | 690 | 1443 |
| **Beamsplitters** | Raman laser detuning | kHz | -34,16 | 2 | 9.2 | 6.17 |
| | Raman laser intensity all pulses | V | 1.8,2.1,2.5 | 0.01 | 219 | 228 |
| | Raman laser intensity one pulse | V | 2.1,2.5,2.9 | 0.01 | -154.5 | 290.9 |
| | Beamsplitter height | ms | 7,11,14 | 0.01 | -26.3 | 17.2 |
| **Interferometer Environment** | $z$ B-field offset | V | -0.6, -0.25, 0. | 0.002 | 91.8 | 450.2 |
| | $x$ B-field offset | V | -0.35, 0.0, 0.25 | 0.002 | 108 | 286 |
| | $y$ B-field offset | V | -0.5, -0.3, -0.1, 0.0, 0.1, 0.2, 0.6 | 0.002 | -53.9 | 208 |
| | MOT B-field applied during interferometer | mG/cm | 15000 | 10 | 0.022 | 0.015 |
| | Tracer intensity | mW | 1,4 | 0.01 | 37.5 | 66.9 |
| | Experiment tilt | V | 3 | 0.05 | 13.8 | 32.4 |
| **Mass Correlated Parameters** | Trap depth correlated with $\widetilde{\mathcal{M}}$ | | 0.2,0.04 | 0.00032 | 340 | 914 |
| | $x$ MOT B-field correlated with $\widetilde{\mathcal{M}}$ | V | 0.5, 0.85 | 0.007 | -57.9 | 259 |
| | $y$ Interferometer B-field correlated with $\widetilde{\mathcal{M}}$ | V | 0.3 | 0.002 | 206 | 413 |
| | $z$ Interferometer B-field correlated with $\widetilde{\mathcal{M}}$ | V | 0.3 | 0.002 | -119 | 404 |
| **Sample prep - after launch** | Velocity selection disabled | | 3 | 0.01 | 54.7 | 121.7 |
| | Velocity selection duration | us | 130, 260 | 1.3 | -1.04 | 0.7 |
| | Velocity selection detuning | kHz | -22, 0, 6, 14 | 2 | 3.587 | 7.22 |
| | Atom number (via microwave $\pi$-pulse duration) | us | 24, 44 | 3 | -9.23 | 6.79 |
| | Launch laser intensity | V | 2,4,8 | 0.1 | 16.3 | 31.8 |
| | Elevator laser intensity | V | 2,4,6,10 | 0.1 | -20.5 | 17.6 |
| **Sample prep - before launch** | RSC duration | ms | 2,4,40 | 0.001 | 1.24 | 2.95 |
| | RSC 1D beam intensity | V | 0.5, 1.1, 2 | 0.1 | -411 | 485 |
| | RSC 2D beam intensity | V | 5, 6, 7, 8, 10 | 0.5 | -43.5 | 37.5 |
| | RSC pumping intensity | arb unit | 1, 0.5 | 0.1 | 164 | 187 |
| | PGC duration | ms | 0, 10, 50 | 0.001 | -21.2 | 18.9 |
| | Hold time after sample prep | ms | 1.8, 200, 500 | 0.001 | 0.055 | 0.4 |
| **Sample prep B-fields** | MOT B-field $x$ offset | V | -0.35, 0.0, 0.55 | 0.05 | -196 | 304 |
| | MOT B-field $y$ offset | V | -0.2, 0.0, 0.2 | 0.05 | 410.2 | 427 |
| | MOT B-field $z$ offset | V | -0.8, -0.6, 0.0 | 0.05 | -102 | 110 |
| **Imaging** | Camera exposure time | ms | 1, 2, 4 | 0.001 | 50.6 | 31.4 |
| | Atom imaging position 2 mm higher | ms | 10 | 0.1 | -3.3 | 13.45 |
| | Atom imaging position 1 mm higher | ms | 5 | 0.1 | -31 | 19 |
| | Blowaway time | ms | 14, 20 | 0.1 | -1.97 | 8.67 |

**Table *S1*. Parameters varied in the search for unexpected systematic errors.** Parameters are categorized by the part of the experimental cycle they belong to. Each parameter is varied over a range that is as large as possible, limited by decreases in signal size or contrast. Slope and uncertainty resulting from fitting the data to a linear slope are shown.